\documentclass[12pt]{article}
\usepackage{pic03}
\usepackage{hyperref}
\usepackage{url}
\usepackage{graphicx}

\begin{document}

\title{\bf CONSTRUCTION AND TESTS OF MODULES FOR THE ATLAS PIXEL DETECTOR}
\author{
J\"{o}rn Grosse-Knetter \\
{\em Physikalisches Institut, Universit\"{a}t Bonn, Nussallee 12, D-53115 Bonn} \\
on behalf of the ATLAS Pixel collaboration}
\maketitle

\baselineskip=14.5pt
\begin{abstract}
The ATLAS Pixel Detector is the innermost layer of the ATLAS tracking 
system and will contribute significantly to  the ATLAS track and vertex 
reconstruction. The detector consists of identical sensor-chip-hybrid modules,
arranged in three barrels in the centre and three disks on either side
for the forward region.

The position of the pixel detector near the interaction point requires 
excellent radiation hardness, mechanical and thermal robustness, good 
long-term stability, all combined with a low material budget. The 
pre-production phase of such pixel modules has nearly finished, yielding 
fully functional modules. Results are presented of tests with these 
modules.\end{abstract}

\baselineskip=17pt

\section{Module Layout}

A pixel module consists of a single n-on-n silicon sensor, 
approx.~2$\times$6~cm$^2$ in size. The sensor is
subdivided into 47,268 pixels which are connected individually to 16 
front-end (FE) chips via "bumps". These chips are connected to a
module-control chip (MCC) mounted on a kapton-flex-hybrid glued 
onto the back-side of the sensor. 
The MCC communicates with the off-detector electronics via 
opto-links, whereas power is fed into the chips via cables connected 
to the flex-hybrid~\cite{general}.

The sensor is subdivided into
41984 ``standard'' pixels of 50~$\mu$m in azimuth times 400~$\mu$m 
parallel to the LHC beam, 
and 5284 ``long'' pixels of $50 \times 600$~$\mu$m
to cover the gaps between adjacent front-end chips. 
The module has 46080 read-out channels, 
which is smaller than the number of pixels because there is a 
200~$\mu$m gap in between
FE chips on opposite sides of the module.
To get full coverage the last eight pixels 
in each row 
are connected to only four channels (``ganged'' pixels).

The connection between each pixel and its read-out channel 
is made through a  bump bond. 
Two technologies are used, indium bumps and solder bumps. 
The minimum bump spacing is 50~$\mu$m. 
No underfill material is used between the bumps to minimise 
the capacitive coupling between pixels 
as well as the capacitive load on the FE inputs. Consequently 
the bumped assembly is mechanically 
held together only by the bumps~\cite{bumps}.

The FE chips contain 2880 individual charge sensitive 
analogue circuits with a digital read-out
that operates at 40~MHz clock. The analogue part consists 
of a high-gain, fast preamplifier followed
by a DC-coupled second stage and a differential discriminator. 
The threshold of the discriminator
ranges up to 1~fC, its nominal value being 0.5~fC. On top of 
a globally set threshold value
the threshold of each of the 2880 channels can be adjusted 
individually to allow fine tuning.
When a hit is detected by the discriminator the 
pixel address and the time at which the hit
occured is provided. Alongside that the time over threshold 
(ToT) information allows reconstruction
of the charge seen by the preamplifier. The pixel address 
and timing/charge information is then passed 
to buffers at the bottom of 
the chip where data waits for a matching trigger
before being sent to the MCC. The MCC does a first 
event-building of the individual data of the
16 FE chips~\cite{MCC}.

\section{Module Tests}% and quality assurance measurements}

In order to assure full functionality of the modules in the 
later experiment, measurements
in a test beam and at the production sites are performed. 

Beam tests are performed at the SPS at CERN 
using 180~GeV/c hadrons. 
The setup consists of a beam telescope for the position 
measurement~\cite{BAT}, trigger 
scintillators for timing measurement to 36~ps, and up to four pixel modules.
The number of defective channels is observed to less than $10^{-3}$ and 
for standard 
pixels the efficiency for normal 
incidence particles is 99.57$\pm$0.15\%. 
The timewalk, i.e. the variation in the time 
when the discriminator input goes above threshold, is an 
issue since hits with a low 
deposited charge have an arrival time later than the ones 
with high charges,
in particular for ganged pixels. This problem
has been addressed in the latest version of FE chips.

\begin{figure}[htbp]
  \centerline{\hbox{ \hspace{0.2cm}
    \includegraphics[width=6.5cm]{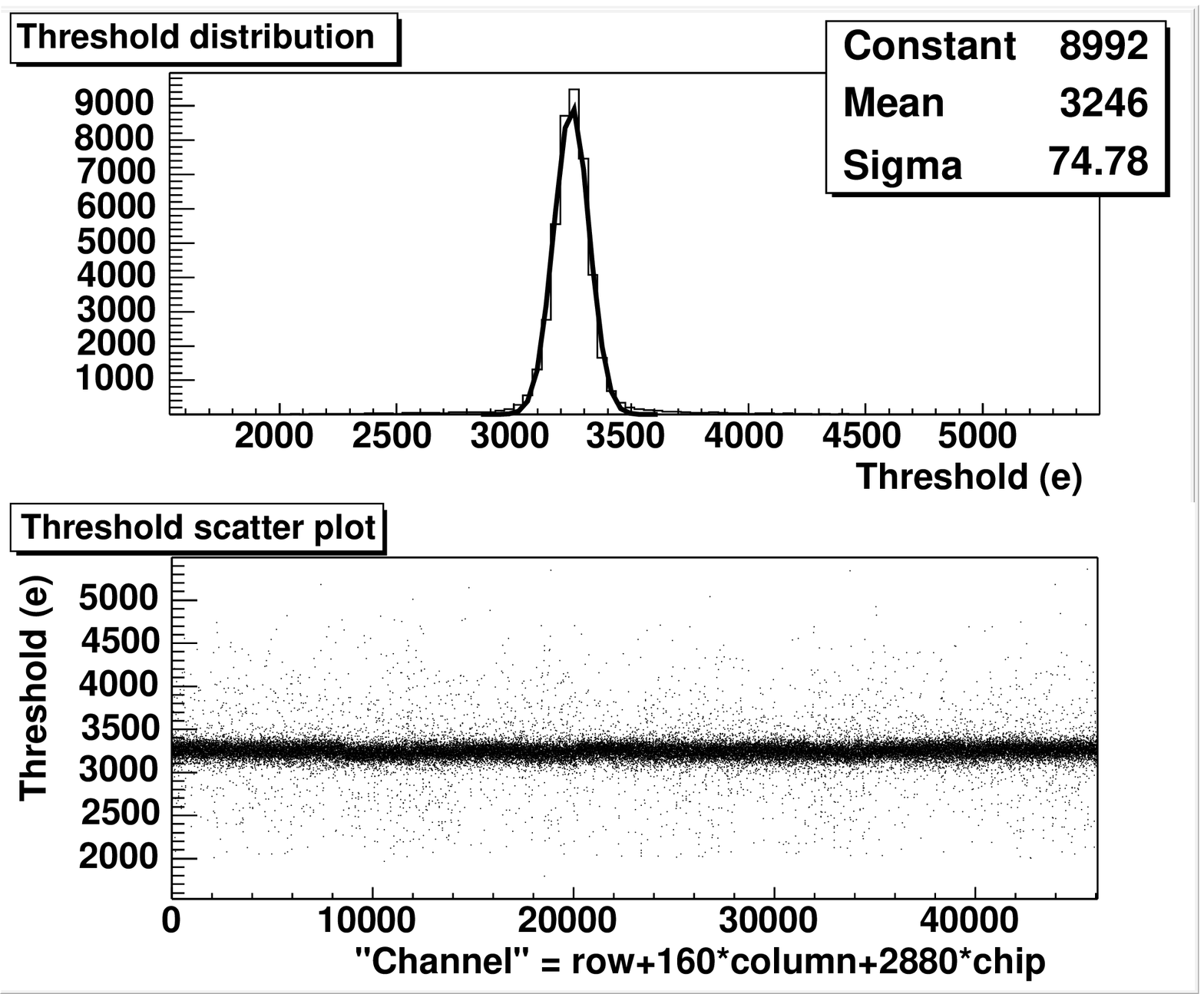}
    \hspace{0.3cm}
    \includegraphics[width=6.5cm]{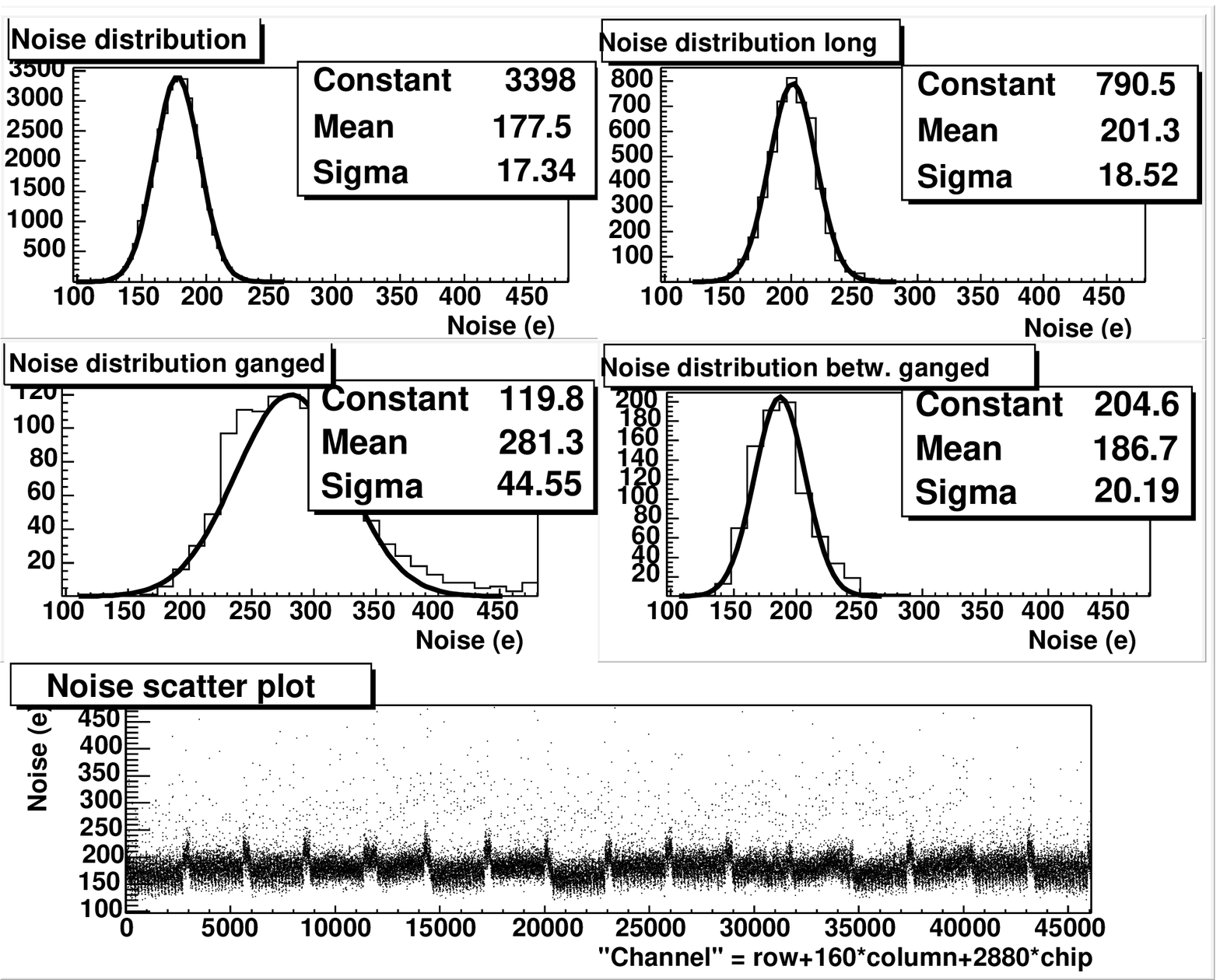}
    }
  }
 \caption{\it
      Threshold (left) and noise (right) distribution of a module
      after individual threshold tuning.
    \label{thrnse} }
\end{figure}

An important test that allows a large range of in-laboratory 
measurements is the threshold scan.
Signals are created with on-module charge injection and 
scanning the number of hits
versus the so injected charge yields the physical value of 
the threshold of the discriminator
and the equivalent noise charge as seen by the preamplifier. 
A set of such scans is used to reduce the
threshold dispersion by adjusting
a parameter individually for each channel (see above). 
The resulting threshold
and noise after threshold tuning is shown in 
figure~\ref{thrnse}.
Typically approx.~100~e threshold dispersion across 
a module and a noise value
of below 200~e for standard pixels is achieved, as is 
needed for good performance.
In a similar fashion, the cross-talk is measured to a few per 
cent for 
standard pixels and a timewalk measurement yields a result similar
to that from the test beam. Data taken when illuminating the 
sensor with a radioactive source
allows in-laboratory detection of defective channels, again 
yielding a number similar
to that of the test beam. The source-spectrum reconstructed 
from the ToT-readings
is in agreement with expectations. 

In addition, modules are irradiated to a dose
approximately corresponding to 10 years of ATLAS operation
The latter results are currently only preliminary and are 
thus not reported here although they look promising.

\end{document}